\documentclass[runningheads]{llncs}

\usepackage{graphicx}
\usepackage{tikz}
\usepackage{xspace}
\usepackage{amssymb}
\usepackage{amsmath}
\usepackage{paralist}
\usepackage{mathtools}
\usepackage{ifpdf}
\usepackage{pseudocode}
\usepackage{chngcntr}

\counterwithout{pseudocode}{section}
\newcounter{newpseudonum}[pseudocode]

\ifpdf
  \providecommand{\refline}[1]{\hyperref[#1]{(\ref*{#1})}}
\else
  \providecommand{\refline}[1]{\ref*{#1}}
\fi

\renewcommand{\ELSE}{\\\pcodetab{1}\mbox{ \bfseries \makebox[0pt][l]{else}\phantom{then} }}
\renewcommand{\RETURN}[1]{\ifthenelse{\equal{#1}{} }{\mbox{\bfseries return}}{\mbox{\bfseries return}#1}}
\newcommand{\FUNCTION}[2]{\mbox{\bfseries proc }\mbox{\textsc{#1}}\left(\ensuremath{#2}\right)\\}

\newlength{\pcodewidth}
\newenvironment{code}[1]{
\begin{Sbox}
\!\!\begin{minipage}{#1}%\pcodewidth}
\bfseries
\noindent
\scriptsize
$$
\begin{array}{@{\hspace*{1ex}}lr@{}}
}{
\end{array}
$$
\end{minipage}\vspace{-2mm}
\end{Sbox}
\shadowbox{\TheSbox}{}
}

\usepackage[xcolor]{changebar}
\usepackage{todonotes}

\presetkeys{todonotes}{color={magenta!80!green!10}, size=\scriptsize, fancyline}{}
\makeatletter
\if@todonotes@disabled

\else

\fi
\makeatother

\newcommand{\finishes}{\ensuremath{\langle E \rangle}\xspace}
\newcommand{\during}{\ensuremath{\langle D \rangle}\xspace}
\newcommand{\starts}{\ensuremath{\langle B \rangle}\xspace}
\newcommand{\overlaps}{\ensuremath{\langle O \rangle}\xspace}
\newcommand{\meets}{\ensuremath{\langle A \rangle}\xspace}

\newcommand{\Toverlaps}{\ensuremath{\langle \overline{O} \rangle}\xspace}

\newcommand{\later}{\ensuremath{\langle L \rangle}\xspace}

\newcommand{\mmodels}{\ensuremath{\Vdash}}

\begin{document}
\title{Interval Temporal Logic Decision Tree Learning}
\author{
Andrea Brunello\inst{2} \orcidID{0000-0003-2063-218X} \and \\
Guido Sciavicco\inst{1} \orcidID{0000-0002-9221-879X} \and \\
Ionel Eduard Stan\inst{2} \orcidID{0000-0001-9260-102X} }
%Third Author\inst{3}\orcidID{2222--3333-4444-5555}}
%
\authorrunning{A. Brunello et al.}
\institute{
		Dept. of Mathematics and Computer Science \\ University of Ferrara (Italy)\\
		\email{guido.sciavicco@unife.it}\and
        Dept. of Mathematics, Computer Science, and Physics \\ University of Udine (Italy) \\
		\email{andrea.brunello@uniud.it} \\
		\email{stan.ioneleduard@spes.uniud.it}
}

\maketitle              % typeset the header of the contribution
\begin{abstract}
Decision trees are simple, yet powerful, classification models used to classify categorical and numerical data, and, despite their simplicity, they are commonly used in operations research and management, as well as in knowledge mining. From a logical point of view, a decision tree can be seen as a structured set of logical rules written in propositional logic. Since knowledge mining is rapidly evolving towards temporal knowledge mining, and since in many cases temporal information is best described by interval temporal logics, propositional logic decision trees may evolve towards interval temporal logic decision trees. In this paper, we define the problem of interval temporal logic decision tree learning, and propose a solution that generalizes classical decision tree learning.

\keywords{Decision trees  \and Interval temporal logics \and Symbolic learning.}
\end{abstract}

\section{Introduction}

It is commonly recognized that modern decision trees are of primary importance among classification models~\cite{witten2016data}. They owe their popularity mainly to the fact that they can be trained and applied efficiently even on big datasets, and that they are easily interpretable, meaning that they are not only useful for prediction per se, but also for understanding the reasons behind the predictions. Interpretability is of extreme importance in domains in which understanding the classification process is as important as the accuracy of the classification itself, such in the case of production business systems or in the computer-aided medicine domain. A typical decision tree is constructed in a recursive manner, following the traditional \emph{Top Down Induction of Decision Trees} (TDIDT) approach~\cite{DBLP:journals/ml/Quinlan86}: starting from the root, at each node the attribute that best partitions the training data, according to a predefined score, is chosen as a test to guide the partitioning of instances into child nodes. The process continues until a sufficiently high degree of purity (with respect to the target class), or a minimum cardinality constraint (with respect to the number of instances reaching the node), is achieved in the generated partitions. This is the case of the well-known decision tree learning algorithm ID3~\cite{DBLP:journals/ml/Quinlan86}, which is the precursor of the commonly-used C4.5~\cite{Quinlan:1987:SDT:50007.50008}. A decision tree can be seen as a structured set of rules: every node of the tree can be thought of as a decision point, and, in this way, each branch becomes a conjunction of such conditional statements, that is, a {\em rule}, whose right-hand part is the class. A conditional statement may have many forms: it can be a yes/no statement (for binary categorical attributes), a categorical value statement (for non-binary categorical attributes), or a splitting value statement (for numerical attributes); the ariety of the resulting tree is two if all attributes are binary or numerical, or more, if there are categorical attributes with more than two categories. Each statement can be equivalently represented with {\em propositional letters}, so that a decision tree can be also seen as a structured set of {\em propositional logic} rules.

\medskip

%\todo[]{Sistemare il fatto che esiste uno spazio tra il cognome di Andrea (guardare in alto la prima riga della pagina) e la virgola}
\noindent{\bf Temporal classification: static solutions.} Just focusing on the static aspects of data is not always adequate for classification; for example, in the medical domain, one may want to take into account which symptoms a patient was experiencing at the same time, or whether two symptoms were overlapping. That is, in some application domains, the temporal aspects of the information may be essential to an accurate prediction. Within static decision tree learning, temporal information may be aggregated in order to circumvent the absence of explicit tools for dealing with temporal information (for example, a patient can be labelled with a natural number describing how many times he/she has been running a fever during the observation period); the ability of a decision tree to perform a precise classification based on such processed data, however, strongly depends on how well data are prepared, and therefore on how well the underlying domain is understood. Alternatively, decision trees have been proposed that use {\em frequent patterns}~\cite{cheng2007discriminative,fournier2014vgen,DBLP:conf/ismis/LinO00}
%~\cite{DBLP:conf/ismis/LinO00,DBLP:books/daglib/0001200,DBLP:books/daglib/0071550}
 in nodes, considering the presence/absence of a frequent pattern as a categorical attribute~\cite{DBLP:conf/icist/BrunelloMMS18,fan2008direct}. Nevertheless, despite being the most common approach to (explicit) temporal data classification, frequent patterns in sequences or series have a limited expressive power, as they are characterized by being {\em existential} and by intrinsically representing temporal information with instantaneous events.

\medskip

\noindent{\bf Our approach: interval temporal logic decision trees.} A different approach to temporal classification is mining temporal logic formulas, and since temporal databases universally adopt an interval-based representation of time, the ideal choice to represent temporal information in data is {\em interval} temporal logic. The most representative propositional interval temporal logic is Halpern and Shoham's Modal Logic of Allen's Relations~\cite{HalpernS91}, also known as HS. Its language encompasses one modal operator for each interval-to-interval relation, such as {\em meets} or {\em before}, and the computational properties of HS and its fragments have been studied in the recent literature (see, e.g.~\cite{sosym,DBLP:journals/tcs/BresolinMMSS14,ijfcs2012}). The very high expressive power of HS, as well as its versatility, make HS the ideal candidate to serve as the basis of a temporal decision tree learning algorithm. Based on these premises, we propose in this paper a decision tree learning algorithm that produces HS-based trees. Our proposal, Temporal ID3, is a direct generalization of the ID3 algorithm~\cite{DBLP:journals/ml/Quinlan86}, founded on the logical interpretation of tree nodes, and focuses on data representation and node generation; we borrow other aspects, such as splitting based on information gain and the overall learning process from the original algorithm. The accuracy of a decision tree and its resilience to over-fitting also depends on the stopping criterion and possible post-pruning operations, but we do not discuss these aspects here.

\medskip

\noindent{\bf Existing approaches to temporal logic decision trees.} Learning temporal logic decision trees is an emerging field in the analysis of physical systems, and, among the most influential approaches, we mention learning of automata~\cite{DBLP:journals/iandc/Angluin87} and learning Signal Temporal Logic (STL) formulas~\cite{DBLP:conf/formats/BartocciBS14,DBLP:conf/isola/BufoBSBLB14,DBLP:conf/hybrid/NguyenKJDBJ17,Rajan:2006:ART:1218776.1218799}. In particular, STL is a point-based temporal logic with {\em until} that encompasses certain metric capabilities, and learning formulas of STL has been focused on both the fine tuning of the metric parameters of a predefined formula and the learning the innermost structure of a formula; among others, decision trees have been used to this end~\cite{DBLP:conf/hybrid/BombaraVPYB16}. Compared with STL decision tree learning, our approach has the advantage of learning formulas written in a well-known, highly expressive interval-based temporal logic language;
%Compared with STL decision tree learning, our approach has the advantage of learning formulas written in a interval-based, instead of point-based, well-known and highly expressive temporal logic language;
because of the nature of the underlying language and of the interval temporal logic models, certain application domains fit naturally into this approach. Moreover, since our solution generalizes the classical decision tree learning algorithm ID3, and, particularly, the notion of information gain, it is not limited to binary classification only. Moreover, in~\cite{Blockeel:1998:TIF:1643275.1643308} a first-order framework for TDIDT is presented with the attempt to make such paradigm more attractive to inductive logic programming (ILP). Such a framework provides a sound basis for logical decision tree induction; in opposition, we employ the framework to represent {\em modal}, instead of first-order, relational data. Additionally, our approach should not be confused with~\cite{Mballo:2005}, in which the term {\em interval} indicates an uncertain numerical value (e.g., {\em the patient has a fever of 38 Celsius} versus {\em the patient has a fever between 37.5 and 38.5 Celsius}), and in which an algorithm for inducing decision trees on such uncertain data is presented that is based on the so-called Kolmogorov-Smirnov criterion, but the data that are object of that study are not necessarily temporal, and the produced trees do not employ any temporal (logical) relation. In~\cite{Antipov2011,8465421} and~\cite{KarimiHamilton}, the authors present two other approaches to a temporal generalization of decision tree learning. In the former, the authors provide a general method for building point-based temporal decision trees, but with no particular emphasis on any supporting formal language. In the latter, the constructed trees can be seen as real-time algorithms that have the ability to make decisions even if the entire description of the instance is not yet available. Finally, in~\cite{DBLP:journals/jair/ConsolePD03}, a generalization of the decision tree model is presented in which nodes are possibly labelled with a timestamp to indicate when a certain condition should be checked.

Summarizing, our approach is essentially different from those presented in the literature in several aspects. As a matter of fact, by giving a logical perspective to decision tree learning, we effectively generalize the learning model to a temporal one, instead of introducing a new paradigm. In this way, instances that present some temporal component are naturally seen as timelines, and, thanks to the expressive power provided by HS, our algorithm can learn a decision tree based on the temporal relations between values, instead of the static information carried by the values.

\section{Preliminaries}\label{sec:prel}

\noindent{\bf Decision trees.} Decision tree induction is based on the following simple concepts~\cite{Quinlan:1987:SDT:50007.50008}. Given a set of observable values $V= \{v_1,v_2,\ldots,v_n\}$, with associated probabilities $\Pi=\{\pi_1,\pi_2,\ldots,\pi_n\}$, the \emph{information conveyed by} $\Pi$ (or \emph{entropy}), is defined as:

$$
E(\Pi) = -\sum_{i=1}^{n} \pi_i\log(\pi_i).
$$

\noindent Assume that a dataset $\mathcal T$ has $n$ instances, each characterized by the attributes $A_1,\ldots,A_l$, and distributed over $s$ classes $C_1,\ldots,C_s$. Each class $C$ can be seen as the subset of $\mathcal T$ composed of precisely those instances classified as $C$, so that the information needed to identify the class of an element of $\mathcal T$ is:

$$
Info(\mathcal T) = E(\{\frac{|\mathcal C_1|}{|\mathcal T|},\frac{|\mathcal C_2|}{|\mathcal T|},\ldots,\frac{|\mathcal C_s|}{|\mathcal T|}\}).
$$

\noindent Intuitively, the entropy is inversely proportional to the purity degree of $\mathcal T$ with respect to the class values. {\em Splitting}, which is the main operation in decision tree learning, is performed over a specific attribute $A$. If $A$ is categorical and its domain $Dom(A)$ consists of $m$ distinct values, we can split $\mathcal T$ into $\mathcal T_1,\ldots,\mathcal T_m$, each $\mathcal T_i$ being characterized by $A$ having precisely the value $a_i$ (i.e., $A=a_i$). The information of a categorical split, therefore, is:

$$
InfoCat(A,\mathcal T) = \sum_{i=1}^m (\frac{|\mathcal T_i|}{|\mathcal T|}) Info(\mathcal T_i).
$$

\noindent If, on the other hand, $A$ is numerical, then the set $\{a_1<\ldots<a_m\}$ of {\em actual} values for $A$ that are present in $\mathcal T$ gives rise to $m-1$ possible splits, all of them binary, and the information conveyed by each possible split is, then, a function not only of the attribute but also of the chosen value:

$$
InfoNum(A,a_i,\mathcal T) = (\frac{|\mathcal T_1|}{|\mathcal T|}) Info(\mathcal T_1)+(\frac{|\mathcal T_2|}{|\mathcal T|}) Info(\mathcal T_2),
$$

\noindent where $\mathcal T_1$ (respectively, $\mathcal T_2$) encompasses all and only those instances with $A\le a_i$ (respectively, $A>a_i$). The information conveyed by an attribute can be consequently defined as:

$$
InfoAtt(A,\mathcal T) = \left\{
\begin{array}{ll}
InfoCat(A,\mathcal T) &\mbox{if $A$ is categorical},\\
\min\limits_{a_i\in Dom(A)} \{InfoNum(A,a_i,\mathcal T)\} &\mbox{if $A$ is numerical},
\end{array}
\right.
$$

\noindent and the {\em information gain} brought by $A$ is defined as:

$$
Gain(A,\mathcal T) = Info(\mathcal T) - InfoAtt(A,\mathcal T).
$$

The information gain, which can be also seen as the reduction of the expected entropy when the attribute $A$ has been chosen, is used to drive the splitting process, that is, to decide over which attribute (and how) the next split is performed. The underlying principle to decision tree building consists of recursively splitting the dataset over the attribute that guarantees the greatest information gain until a certain stopping criterion is met. Each split can be seen as a propositional condition {\em if $p$ then -, else -}. When splitting is performed over a numerical attribute, e.g., $A\le a_i$, then the corresponding propositional statement is simply the condition itself (in our example, is a propositional letter $p_{A\le a_i}$); when it is performed over a categorical attribute, e.g., $A=a_1$, $A=a_2$, \ldots, then each statement is a propositional statement (in our example, $p_{A=a_1}$, $p_{A=a_2}$,\ldots) on its own.

\medskip

\begin{figure*}[t]
\centering
\begin{tikzpicture}[scale=.75]%[scale=.58]
%  \tikzstyle{every node}=[font=\tiny]
 \tikzstyle{every node}=[font=\footnotesize]
% \tikzstyle{every node}=[font=\scriptsize]
%\tikzstyle{every node}=[font=\scriptsize]
% \draw (0,0)node(op){\hs \bf operators};

\draw (0,0)node(op){\bf HS};

% \draw (4,0)node{\bf Corresponding relations};
%
% \draw (op) ++(-1.5,-.25) -- ++(12,0);

\draw (op) ++(0,-1.5)node(meets){$\meets$};
\draw (meets)++(0,-.75)node(later){$\later$};
\draw (meets)++(0,-1.5)node(starts){$\starts$};
\draw (meets)++(0,-2.25)node(finishes){$\finishes$};
\draw (meets)++(0,-3)node(during){$\during$};
\draw (meets)++(0,-3.75)node(overlaps){$\overlaps$};

%\draw (meets)++(0,-5)node(op){\bf \hsthr/\hssev};

\draw (meets)++(1.7,0) node(sep1){} ++(0,.5) -- ++(0,-4.7);

%\draw (meets)++(1.7,0) node(sep4){} ++(0,-5.5) -- ++(0,-2.75);
%\draw (op) ++(0,-1.25)node(meetsov){$\langle AO\rangle$};
%\draw (op) ++(0,-2)node(dbe){$\langle DBE\rangle$};
%\draw (op) ++(0,-2.75)node(i){$\langle I\rangle$};

\draw (sep1)++(3,1.5)node(rel){\bf Allen's relations};
%\draw (rel)++(0,-6.3)node(sem){\bf Semantics};

\draw (sep1)++(0.5,0)node[right](Ra){$[x,y] R_A [x',y'] \Leftrightarrow y=x'$};
\draw (sep1)++(0.5,-.75)node[right](Rl){$[x,y] R_L [x',y'] \Leftrightarrow y < x'$};
\draw (sep1)++(0.5,-1.5)node[right](Rs){$[x,y] R_B [x',y'] \Leftrightarrow x=x', y' < y$};
\draw (sep1)++(0.5,-2.25)node[right](Rf){$[x,y] R_E [x',y'] \Leftrightarrow y=y', x < x'$};
\draw (sep1)++(0.5,-3)node[right](Rd){$[x,y] R_D [x',y'] \Leftrightarrow x < x', y' < y$};
\draw (sep1)++(0.5,-3.75)node[right](Ro){$[x,y] R_O [x',y'] \Leftrightarrow x < x' < y < y'$};
% \draw (meets)++(.7,-3.75)node[right](Ro){$[a,b] \textbf{$R$} [c,d]$ iff $a < c < b < d$};

%\draw (sep1)++(0.5,-6.3)node[right](Rao){$\langle AO\rangle\equiv \meets\vee\overlaps$};
%\draw (sep1)++(0.5,-7.05)node[right](Rdbe){$\langle DBE\rangle\equiv \during\vee\starts\vee\finishes$};
%\draw (sep1)++(0.5,-7.8)node[right](Ri){$\langle I\rangle\equiv \meets\vee\Tmeets\vee\overlaps\vee\Toverlaps\vee\langle DBE\rangle\vee\langle\overline{DBE}\rangle$};

\draw (sep1)++(8.5,0)node(sep2){} ++(0,0.5) -- ++(0,-4.7);

\draw (sep2)++(2.8,1.5)node(graphic){\bf Graphical representation};

\draw[red,|-|] (sep2) ++(.7,.75)node[above](a){\tiny $x$} -- ++(2,0)node[above](b){\tiny $y$};
\draw[dashed,red,help lines,thick] (a) -- ++(0,-5.25);
\draw[dashed,red,help lines,thick] (b) -- ++(0,-5.25);

\draw[|-|] (b) ++(0,-1) ++(0,0)node[above](Ac){\tiny $x'$}
-- ++(1,0)node[above](Ad){\tiny $y'$};
\draw[|-|] (b) ++(0,-1) ++(.5,-.75)node[above](Lc){\tiny $x'$}
-- ++(1,0)node[above](Ld){\tiny $y'$};
\draw[|-|] (a) ++(0,-1) ++(0,-1.5)node[above](Bc){\tiny $x'$}
-- ++(.5,0)node[above](Bd){\tiny $y'$};
\draw[|-|] (b) ++(0,-1) ++(-.5,-2.25)node[above](Ec){\tiny $x'$}
-- ++(.5,0)node[above](Ed){\tiny $y'$};
\draw[|-|] (a) ++(0,-1) ++(.5,-3)node[above](Dc){\tiny $x'$}
-- ++(1,0)node[above](Dd){\tiny $y'$};
\draw[|-|] (a) ++(0,-1) ++(1,-3.75)node[above](Oc){\tiny $x'$}
-- ++(2,0)node[above](Od){\tiny $y'$};
% $[a,b] R_A [c,d]$ & iff $b=c$ \tikz\draw[|-|] (0,0) -- ++(1,0); \\
\end{tikzpicture}

% \vspace{-1mm}

\caption{Allen's interval relations and HS modalities.}
\label{fig:relations}
\end{figure*}

\noindent {\bf Interval temporal logic}. Let $\mathbb D\subseteq\mathbb N$. In the {\em strict} interpretation, an {\em interval} over $\mathbb D$ is an ordered pair $[x,y]$, where $x,y \in \mathbb D$ and $x < y$, and we denote by $\mathbb I(\mathbb D)$ the set of all intervals over $\mathbb D$. If we exclude the identity relation, there are 12 different Allen's relations between two intervals in a linear order~\cite{allen83}: the six relations $R_A$
(adjacent to), $R_L$ (later than), $R_B$ (begins), $R_E$ (ends), $R_D$ (during), and
$R_O$ (overlaps), depicted in Fig.~\ref{fig:relations}, and their inverses, that is, $R_{\overline{X}}
= (R_{X})^{-1}$, for each $X \in \mathcal A$, where $\mathcal A=\{ A, L, B, E, D, O \}$. Halpern and Shoham's modal logic of temporal intervals (HS) is defined from a set of propositional letters $\mathcal{AP}$, and by associating a universal modality $[X]$ and an existential one $\langle X\rangle$ to each Allen's relation $R_{X}$. Formulas of HS are obtained by

$$
\varphi ::= p \mid \neg \varphi \mid \varphi \vee \varphi \mid \langle X \rangle
\varphi\mid \langle \overline X \rangle
\varphi,
$$

\noindent where $p \in \mathcal{AP}$ and $X \in \mathcal A$. The other Boolean connectives and the logical constants, e.g., $\rightarrow$ and $\top$, as well as the universal modalities $[X]$, can be defined in the standard way. For each $X \in \mathcal A$, the modality $\langle \overline{X} \rangle$ (corresponding to the inverse relation $R_{\overline{X}}$ of $R_{X}$) is said to be the {\em transpose} of the modalities $\langle X \rangle$, and vice versa. The semantics of HS formulas is given in terms of {\em timelines} $T
= \langle \mathbb I(\mathbb D),V\rangle$\footnote{We deliberately use the symbol $T$ to indicate both a timeline and an instance in a dataset.}, where $\mathbb D$ is a linear order and $V : \mathcal{AP} \rightarrow 2^{\mathbb{I}(\mathbb D)}$ is a \emph{valuation function}
which assigns to each atomic proposition $p \in \mathcal{AP}$ the set of intervals $V(p)$
on which $p$ holds. The {\em truth} of a formula $\varphi$ on a given interval $[x,y]$ in an interval model $T$ is defined by structural induction on formulas as follows:

\begin{equation*}
\begin{array}{lll}
T,[x,y]\mmodels p& \mbox{if}&[x,y]\in V(p),\mbox{ for }p\in\mathcal{AP};\\
T,[x,y]\mmodels\neg\psi&\mbox{if}&T,[x,y]\not\mmodels\psi;\\
T,[x,y]\mmodels\psi\vee\xi&\mbox{if}&T,[x,y]\mmodels\psi\mbox{ or }T,[x,y]\mmodels\xi;\\
T,[x,y]\mmodels\langle X\rangle\psi&\mbox{if}& \mbox{there is }[w,z]\mbox{ s.t }[x,y] R_{X} [w,z]\mbox{ and }T,[w,z]\mmodels\psi;\\
T,[x,y]\mmodels\langle \bar X\rangle\psi&\mbox{if}& \mbox{there is }[w,z]\mbox{ s.t }[x,y] R_{\bar X} [w,z]\mbox{ and }T,[w,z]\mmodels\psi.
\end{array}
\end{equation*}

HS is a very general interval temporal language and its satisfiability problem is undecidable~\cite{HalpernS91}. Our purpose here, however, is to study the problem of formula {\em induction} in the form of decision trees, and not of formula {\em deduction}, and therefore the computational properties of the satisfiability problem can be ignored at this stage.

\section{Motivations}\label{sec:mot}

\begin{figure}[t]
\begin{center}
\begin{tikzpicture}[scale=.80]
\node(original1) at (4,1) {
\begin{tabular}{|l|l|l|}
\hline
Patient & Symptom & TimeStamp  \\
\hline
%febbri
$P1$ & $fever$ & [3,4] \\
$P2$ & $fever$ & [4,5] \\
$P3$ & $fever$ & [3,5] \\
%mal di testa
$P1$ & $head$ & [2,4] \\
$P2$ & $head$ & [3,5] \\
$P3$ & $head$ & [2,4] \\
$P4$ & $head$ & [4,6] \\
\hline
\end{tabular}
};
\node(original2) at (9,1) {
\begin{tabular}{|l|l|}
\hline
Patient & Class \\
\hline
%febbri
$P1$ & $C1$ \\
$P2$ & $C1$ \\
$P3$ & $C2$ \\
$P4$ & $C2$ \\
\hline
\end{tabular}
};
\node(static) at (1,-5) {
\begin{tabular}{|l|l|l|l|}
\hline
Patient & $fever$ & $head$ & Class \\
\hline
$P1$ & $yes$ & $yes$ & $C1$\\
$P2$ & $yes$ & $yes$ & $C1$\\
$P3$ & $yes$ & $yes$ & $C2$\\
$P4$ & $no$  & $yes$ & $C2$\\
\hline
\end{tabular}
};
\foreach \x in {7, 8, 9, 10, 11, 12, 13}
    \draw (\x cm, -39.5mm) -- (\x cm, -40.5mm);
\foreach \x in {7, 8, 9, 10, 11, 12, 13}
    \draw (\x cm, -49.5mm) -- (\x cm, -50.5mm);
\foreach \x in {7, 8, 9, 10, 11, 12, 13}
    \draw (\x cm, -59.5mm) -- (\x cm, -60.5mm);
\foreach \x in {7, 8, 9, 10, 11, 12, 13}
    \draw (\x cm, -69.5mm) -- (\x cm, -70.5mm);
\draw[black, opacity=0.5,thin]  (7,-4) -- (13,-4);
\draw[black, opacity=0.5,thin]  (7,-5) -- (13,-5);
\draw[black, opacity=0.5,thin]  (7,-6) -- (13,-6);
\draw[black, opacity=0.5,thin]  (7,-7) -- (13,-7);
\node(p1t) at (6.5,-4){$P1$};
\node(p2t) at (6.5,-5){$P2$};
\node(p3t) at (6.5,-6){$P3$};
\node(p4t) at (6.5,-7){$P4$};
%eventi
\draw[black, opacity=0.5,thin]  (10,-3.75) -- (11,-3.75);
\node[draw=none] (fever1) at (10.5,-3.65) {\begin{tiny}$fever$\end{tiny}};
\draw[black, opacity=0.5,thin]  (9,-3.5) -- (11,-3.5);
\node[draw=none] (head1) at (10,-3.37) {\begin{tiny}$head$\end{tiny}};

\draw[black, opacity=0.5,thin]  (11,-4.75) -- (12,-4.75);
\node[draw=none] (fever2) at (11.5,-4.65) {\begin{tiny}$fever$\end{tiny}};
\draw[black, opacity=0.5,thin]  (10,-4.5) -- (12,-4.5);
\node[draw=none] (head2) at (11,-4.37) {\begin{tiny}$head$\end{tiny}};

\draw[black, opacity=0.5,thin]  (10,-5.75) -- (12,-5.75);
\node[draw=none] (fever3) at (11,-5.65) {\begin{tiny}$fever$\end{tiny}};
\draw[black, opacity=0.5,thin]  (9,-5.5) -- (11,-5.5);
\node[draw=none] (head3) at (10,-5.37) {\begin{tiny}$head$\end{tiny}};

\draw[black, opacity=0.5,thin]  (11,-6.5) -- (13,-6.5);
\node[draw=none] (head1) at (12,-6.37) {\begin{tiny}$head$\end{tiny}};

\node(p1t) at (13.45,-4){$C1$};
\node(p2t) at (13.45,-5){$C1$};
\node(p3t) at (13.45,-6){$C2$};
\node(p4t) at (13.45,-7){$C2$};

\node(p1t) at (7, -7.5){\tiny$0$};
\node(p1t) at (8, -7.5){\tiny$1$};
\node(p1t) at (9, -7.5){\tiny$2$};
\node(p1t) at (10, -7.5){\tiny$3$};
\node(p1t) at (11, -7.5){\tiny$4$};
\node(p1t) at (12, -7.5){\tiny$5$};
\node(p1t) at (13, -7.5){\tiny$6$};

\node[draw=none](middle) at (6,-1.5) {};
\node[draw=none](sx) at (3,-2.5) {};
\node[draw=none](dx) at (9,-2.5) {};

\path[->] (middle) edge node[sloped,above] {static} (sx);
\path[->] (middle) edge node[sloped,above] {temporal} (dx);
%\draw[->,](middle) -- (sx);
%\draw[->](middle) -- (dx);
\end{tikzpicture}
\end{center} 
\caption{Example of static and temporal treatment of information in the medical domain.}\label{fig:example}
\end{figure}
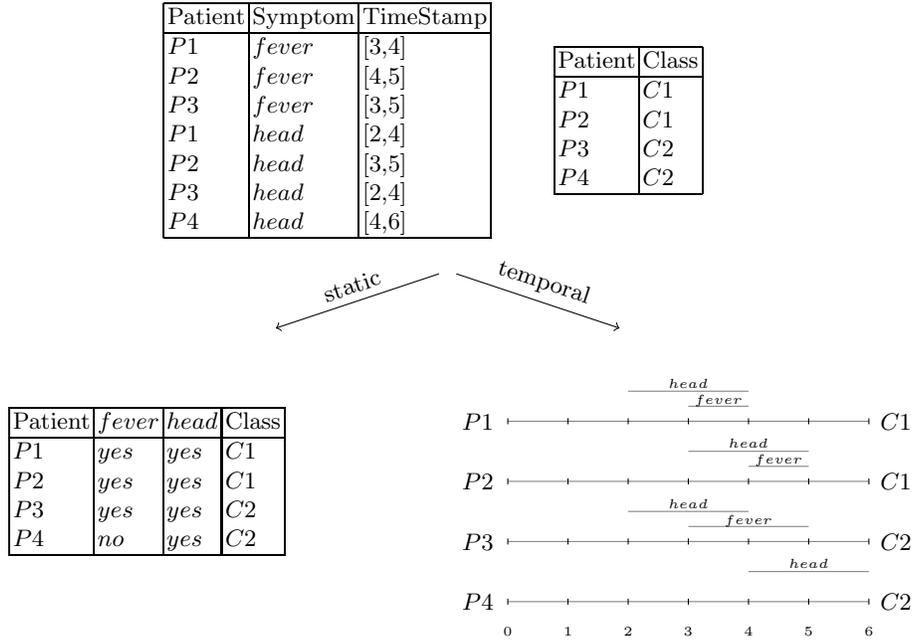

In this section, we present some realistic scenarios in which learning a temporal decision tree may be convenient, and, then, we discuss aspects of data preprocessing related to the temporal component.

\medskip

\noindent{\bf Learning.} There are several application domains in which learning a temporal decision tree may be useful. Consider, for example, a medical scenario in which we consider a dataset of classified patients, each one characterized by its medical history, as in Fig.~\ref{fig:example}, top. Assume, first, that we are interested in learning a {\em static} (propositional) classification model. The history of our patients, that is, the collection of all relevant pieces of information about tests, results, symptoms, and hospitalizations of the patient that occurred during the entire observation period, must be processed so that temporal information is subsumed in propositional letters. For instance, if some patient has been running a fever during the observation period, we may use a proposition $fever$, with positive values for those patient that have had fever, and negative values for the others (as in Fig.~\ref{fig:example}, bottom, left). Depending on the specific case, we may, instead, use the actual temperature of each patient, and a static decision tree learning system may split over $fever<t$, for some threshold temperature $t$, effectively introducing a new propositional letter, and therefore a binary split. Either way, the temporal information is lost in the preprocessing. For example, we can no longer take into account whether $fever$ occurred before, after, or while the patient was experiencing headache ($head$), which may be a relevant information for a classification model. By generating, instead, the timeline of each patient (as in Fig.~\ref{fig:example}, bottom, right), we keep all events and their relative qualitative relations. By learning a decision tree on a preprocessed dataset such as the one in Fig.~\ref{fig:example} (bottom, left), we see that the attribute $head$ has zero variance, and therefore zero predictive capabilities; then, we are forced to build a decision tree using attribute $fever$ alone, which results in a classifier with 75\% accuracy. On the contrary, by using the temporal information in the learning process, we are able to distinguish the two classes: $C1$ is characterized by presenting both $head$ and $fever$, but not overlapping, and this classifier has, in this toy example, $100\%$ accuracy. In this example, the term {\em accuracy} refers to the {\em training set accuracy} (we do not consider independent trainining and test data), that is, the ability of the classification system to discern among classes on the data used to train the system itself; it should not be confused with {\em test set accuracy}, which measures the real classification performances that can be expected on future, real-life examples.

% \medskip

Alternatively, consider a problem in the natural language processing domain. In this scenario, a timeline may represent a {\em conversation} between two individuals. It is known that, in automatic processing of conversations, it is sometimes interesting to label each interval of time with one or more {\em context}, that is, a particular topic that is being discussed~\cite{DBLP:conf/iwcs/Alluhaibi15,DBLP:conf/cicling/Baeza-Yates04,pratt}, in order to discover the existence of unexpected or interesting temporal relations among them. Suppose, for example, that a certain company wants to analyze conversations between selling agents and potential customers: the agents contact the customers with the aim of selling a certain product, and it is known that certain contexts, such as the price of the product ({\em price}), its known advantages ({\em advantages}) over other products, and its possible minor defects ({\em disadvantages}) are interesting. Assume that each conversation has been previously classified between those that have been successful and those that ended without the product being acquired. Now, we want to learn a model able to predict such an outcome. By using only static information, nearly every conversation would be labelled with the three contexts, effectively hiding the underlying knowledge, if it exists. By keeping the relative temporal relations between contexts, instead, we may learn useful information, such as, for example, {\em if price and disadvantages are not discussed together, the conversation will be likely successful}.

\medskip

\noindent{\bf Preprocessing.} Observe, now, how switching from static to temporal information influences data preparation. First, in a context such as the one described in our first example, numerical attributes may become less interesting: for instance, the information on {\em how many times} a certain symptom occurred, or its {\em frequency}, are not needed anymore, considering that each occurrence is taken into account in the timeline. Moreover, since the focus is on attributes {\em relative temporal positions}, even categorical attributes may be ignored in some contexts: for instance, in our scenario, we may be interested in establishing the predictive value of the relative temporal position of $fever$ and $head$ regardless of the sex or age of the patient. It is also worth underlying that propositional attributes over intervals allow us to express a variety of situations, and sometimes propositional labelling may result in {\em gaining} information, instead of loosing it. Consider, again, the case of fever, and suppose that a certain patient is experiencing low fever in an interval $[x,y]$, say, a given day, and that during just one hour of that day, that is, over the interval $[w,z]$ strictly contained in $[x,y]$, he/she has an episode of high fever. A natural choice is to represent such a situation by labelling the interval $[x,y]$ with $lo$ and its sub-interval $[w,z]$ with $hi$.  On the other hand, representing the same pieces of information as three intervals $[x,w], [w,z],[z,y]$ respectively labelled with $lo$, $hi$, and $lo$, which would be the case with a point-based representation (or with an interval-based representation under the homogeneity assumption), would be unnatural, and it would entail hiding a potentially important information such as: ``{\em the patient presented low fever  during the entire day, except for a brief episode of high fever}''. Building on such considerations, our approach in the rest of this paper is based on propositional, non-numerical attributes only.

%In some applicative contexts, however, some numerical attributes cannot be completely ignored. Suppose, for instance, that in our driving medical scenario one is interested in learning how the actual body temperature, in combination with experiencing a headache, influences the class. In this situation, preprocessing data by using the label {\em fever} may not be adequate. To circumvent this problem, one may introduce the letters $fever_{>t}$ for $t$ ranging in set of interesting values (for example, from 36 to 43 Celsius degrees). Data should be preprocessed in such a way that the following simple universal requirement is met:
%
%$$\forall t(fever_{>t}\rightarrow fever_{>(t-1)}).$$
%
%\noindent In this way, the propositional characterization of timelines is recovered, and a temporal decision tree can be learned that effectively takes into account the quantitative value, if it is indeed interesting.

\section{Learning Interval Temporal Logic Decision Trees}\label{sec:algo}

In this section we describe a generalization of the algorithm ID3 that is capable of learning a binary decision tree over a temporal dataset, as in the examples of the previous section; as in classical decision trees, every branch of a temporal decision tree can be read as a logical formula, but instead of classical propositional logic we use the temporal logic HS. To this end, we generalize the notion of information gain, while, at this stage, we do not discuss pre-pruning, post-pruning, and purity degree of a sub-tree~\cite{cart84,Quinlan:1987:SDT:50007.50008}.

\medskip

%\begin{figure}[t]
%\centering
%\begin{tabular}{p{8cm}}
%$p: [0,1],[0,2],[0,3],[1,3],[2,3],\ldots$\\
%$q: [0,1],[0,3],\ldots$ \\
%\ldots  \\
%\end{tabular}
%\caption{A succinct representation of a timeline: each line encodes the valuation function $V$ for a specific proposition.}
%\label{fig:rep}
%\end{figure}

\noindent{\bf Data preparation and presentation.} We assume that the input dataset contains timelines as instances. For the sake of simplicity, we also assume that all timelines are based on the same finite domain $\mathbb D$ of length $N$ (from 0 to $N-1$). The dataset $\mathcal T$ can be seen as an array of $n$ structures; $\mathcal T[j]$ represents the $j$-th timeline of the dataset, and it can be thought of as an interval model. Given a dataset $\mathcal T$, we denote by $\mathcal{AP}$ the set of all propositional letters that occur in it.

\medskip

\noindent{\bf Temporal information.} We are going to design the learning process based on the same principles of classical decision tree learning. This means that we need to define a notion of splitting as well as a notion of information conveyed by a split, and, to this end, we shall use the truth relation as defined in Section~\ref{sec:prel} applied to a timeline. Unlike the atemporal case, splits are not performed over attributes, but, instead, over propositional letters. Splitting is defined {\em relatively to an interval} $[x,y]$, and it can be local, if it is applied on $[x,y]$ itself, or temporal, in which case it depends on the existence of an interval $[z,t]$ related to $[x,y]$ and the particular relation $R_X$ such that $[x,y]R_X[z,t]$ (or the other way around). A {\em local split} of $\mathcal T$ into $\mathcal T_1$ and $\mathcal T_2$, where $[x,y]$ is the \emph{reference} interval of $\mathcal T$, and $p$ is the propositional letter over which the split takes place is defined by:

\begin{equation}\label{localsplit}
\begin{array}{lll}
\mathcal T_1 & = & \{T\in \mathcal T\mid T,[x,y]\mmodels p\},\\
\mathcal T_2 & = & \{T\in \mathcal T\mid T,[x,y]\mmodels \neg p\}.\\
\end{array}
\end{equation}

\noindent On the contrary, a {\em temporal split}, in the same situation, over the temporal relation $R_X$, is defined by:

\begin{equation}\label{temporalsplit}
\begin{array}{lll}
\mathcal T_1 & = & \{T\in \mathcal T\mid T,[x,y]\mmodels\langle X\rangle p\},\\
\mathcal T_2 & = & \{T\in \mathcal T\mid T,[x,y]\mmodels[X]\neg p\}.\\
\end{array}\end{equation}

\noindent Consequently, the {\em local information gain} of a propositional letter $p$ is defined as:

$$LocalGain(p,\mathcal T)=Info(\mathcal T)-\left((\frac{|\mathcal T_1|}{|\mathcal T|}) Info(\mathcal T_1)+(\frac{|\mathcal T_2|}{|\mathcal T|}) Info(\mathcal T_2)\right),$$

\noindent where $\mathcal T_1$ and $\mathcal T_2$ are defined as in (\ref{localsplit}), while the {\em temporal information gain} of a propositional letter $p$ is defined as:

$$TemporalGain(p,\mathcal T)=Info(\mathcal T)-\min_{X\in\mathcal A}\left\{(\frac{|\mathcal T_1|}{|\mathcal T|}) Info(\mathcal T_1)+(\frac{|\mathcal T_2|}{|\mathcal T|}) Info(\mathcal T_2)\right\},$$

\noindent where $\mathcal T_1$ and $\mathcal T_2$ are defined as in (\ref{temporalsplit}) and depend on the relation $R_X$. Therefore, the information gain of a propositional letter becomes:

$$
Gain(p,\mathcal T)=\max\{LocalGain(p,\mathcal T),TemporalGain(p,\mathcal T)\}, 
$$

%
%
%
%Moreover, while in modal logic the equality relation does not play any role, we need to consider possible splits over the current interval. To this end, and with the purpose of keeping the notation as simple as possible, we consider the set $\mathcal A$ of temporal relations defined in Section~\ref{sec:prel}, we define the set $\mathcal A^*$ that extends $\mathcal A$ with the equality relation, and, assuming that each timeline $T\in\mathcal T$ is associated with its current reference interval $[x,y]$, we define:
%
%$$
%InfoTemp(p,X,\mathcal T) = (\frac{|\mathcal T_1|}{|\mathcal T|}) Info(\mathcal T_1)+(\frac{|\mathcal T_2|}{|\mathcal T|}) Info(\mathcal T_2),%$
%$$
%
%\noindent where $p\in\mathcal{AP}$ and $X \in \mathcal A^*$, and:
%$$
%\begin{array}{lll}
%\mathcal T_1 & = & \{T\in \mathcal T\mid T,[x,y]\mmodels\langle X\rangle p\},\\
%\mathcal T_2 & = & \{T\in \mathcal T\mid T,[x,y]\mmodels[X]\neg p\}.\\
%\end{array}
% $$
%
%\noindent Consequently, the information gain conveyed by $p\in\mathcal{AP}$ and $X\in\mathcal A^*$ is:
%
%$$
%Gain(p,X,\mathcal T)=Info(\mathcal T)-InfoTemp(p,X,\mathcal T),
%$$

\noindent and, at each step, we aim to find the letter that maximizes the gain.

\medskip

\medskip

%\begin{algorithm}[t]
%\caption{Temporal ID3}\label{algo}
%\begin{algorithmic}[1]
%\REQUIRE $\mathcal T$
%\REQUIRE the current node $c$
%\IF{a stopping condition is met}
%\RETURN $c$;
%\ENDIF
%\STATE {find $p\in\mathcal{AP},X\in \mathcal A$ such that $Gain(p,\mathcal T)$ is maximal};
%\STATE {compute a new reference interval for each $T\in\mathcal T_1$};
%\STATE {create two children $c_1$ and $c_2$};
%\STATE {call Temporal ID3 on $\mathcal T_1,c_1$};
%\STATE {call Temporal ID3 on $\mathcal T_2,c_2$};
%\RETURN
%\end{algorithmic}
%\end{algorithm}

%\begin{figure}[t]
%%\scriptsize
%\centering
%\begin{code}{120pt}
%\FUNCTION{TemporalID3}{\mathcal T}
%\BEGIN
%AssignBestInterval(\mathcal T)\\
%Learn(\mathcal T,null)\\
%\END\\
%\end{code}
%\begin{code}{140pt}
%\FUNCTION{Learn}{\mathcal T,c}
%\BEGIN
%\IF\ StopCriterion(\mathcal T)=true \THEN
%\RETURN\ c\\
%g=0\\
%\FOR (p\in\mathcal{AP}\AND X\in\mathcal A)\\
%\BEGIN
%\IF (Gain(p,X,\mathcal T)>g) \THEN\\
%\BEGIN
%\widetilde{p}=p\\
%\widetilde{X}=X\\
%g=Gain(p,X,\mathcal T)\\
%\END\\
%\END\\
%(\mathcal T_1,\mathcal T_2)=Split(\mathcal T,p,X)\\
%AssignNewInterval(T_1,p,X)\\
%c_1=CreateLeftChild(c)\\
%c_2=CreateRightChild(c)\\
%Learn(\mathcal T_1,c_1)\\
%AssignBestInterval(\mathcal T_2)\\
%Learn(\mathcal T_2,c_2)\\
%\END\\
%\end{code}
%\caption{The algorithm Temporal ID3.}\label{algo}
%\end{figure}

\newcommand{\squaredpr}{[}

\begin{figure}[t]
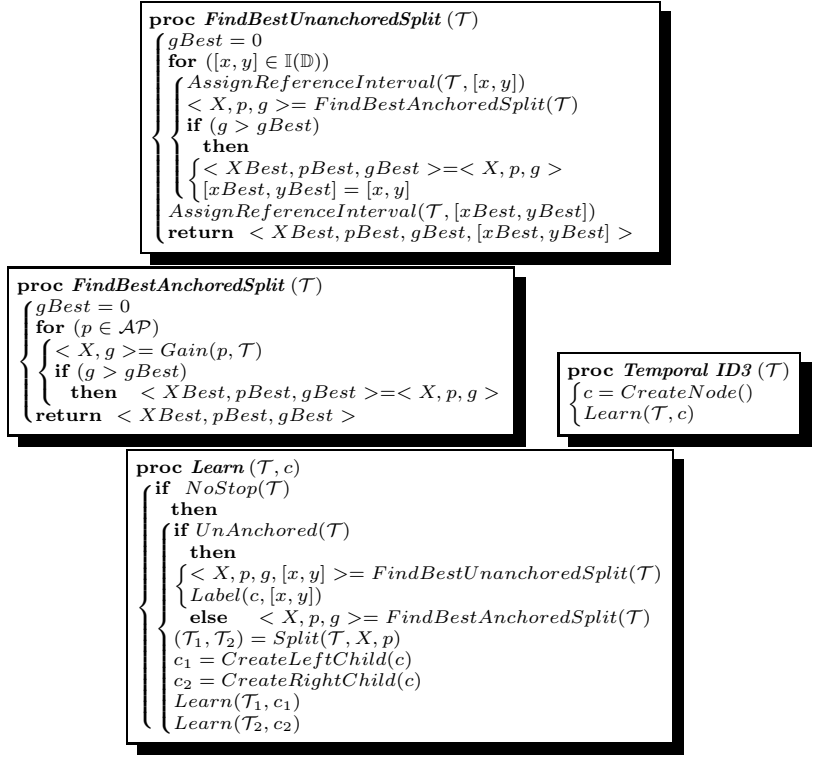

%\scriptsize
\centering
\begin{code}{190pt}
\FUNCTION{{\em FindBestUnanchoredSplit}}{\mathcal T}
\BEGIN
 gBest=0\\
\FOR ([x,y]\in\mathbb{I(D)}) \\
\BEGIN
AssignReferenceInterval(\mathcal T,[x,y])\\
<X,p,g>=FindBestAnchoredSplit(\mathcal T)\\
\IF (g>gBest)\THEN\\
\BEGIN
  <XBest,pBest,gBest>=<X,p,g>\\
  \squaredpr xBest,yBest]=[x,y]\\
\END\\
\END\\
AssignReferenceInterval(\mathcal T,[xBest,yBest])\\
\RETURN\ <XBest,pBest,gBest,[xBest,yBest]>\\
\END\\
\end{code}

\begin{code}{185pt}
\FUNCTION{{\em FindBestAnchoredSplit}}{\mathcal T}
\BEGIN
gBest=0\\
\FOR (p\in\mathcal{AP})\\
\BEGIN
<X,g>=Gain(p,\mathcal T)\\
\IF (g>gBest) \THEN\ <XBest,pBest,gBest>=<X,p,g>\\
\END\\
\RETURN\ <XBest,pBest,gBest>\\
\END\\
\end{code}
\begin{code}{85pt}
\FUNCTION{{\em Temporal ID3}}{\mathcal T}
\BEGIN
c=CreateNode()\\
Learn(\mathcal T,c)\\
\END
\end{code}
\begin{code}{200pt}
\FUNCTION{{\em Learn}}{\mathcal T,c}
\BEGIN
\IF\ NoStop(\mathcal T) \THEN\\
\BEGIN
\IF UnAnchored(\mathcal T) \THEN\\
\BEGIN
<X,p,g,[x,y]>=FindBestUnanchoredSplit(\mathcal T)\\
Label(c,[x,y])\\
\END
\ELSE\ <X,p,g>=FindBestAnchoredSplit(\mathcal T)\\
(\mathcal T_1,\mathcal T_2)=Split(\mathcal T,X,p)\\
c_1=CreateLeftChild(c)\\
c_2=CreateRightChild(c)\\
Learn(\mathcal T_1,c_1)\\
Learn(\mathcal T_2,c_2)\\
\END\\
\END
\end{code}
\caption{The algorithm Temporal ID3.}\label{algo}
\end{figure}

\noindent{\bf The algorithm.} Let us analyze the code in Fig.~\ref{algo}. At the beginning, the timelines in $\mathcal T$ are not assigned any reference interval, and we say that the dataset is {\em unanchored}. The procedure  {\em FindBestUnanchoredSplit} systematically explores every possible reference interval of an unanchored dataset, and, for each one of them, calls {\em FindBestAnchoredSplit}, which, in turn, tries every propositional letter (and, implicitly, every temporal relation) in the search of the best split. This procedure returns the best possible triple $<X,p,g>$, where $X$ is an interval relation, if the best split is temporal, or it has no value, if the best split is local, $p$ is a propositional letter, and $g$ is the information gain. {\em Temporal ID3} first creates a root node, and then calls {\em Learn}. The latter, in turn, first checks possible stopping conditions, and then finds the best split into two datasets $\mathcal T_1$ and $\mathcal T_2$. Of these, the former is now {\em anchored} (to the reference interval returned by  {\em FindBestUnanchoredSplit}), while the latter is still unanchored. During a recursive call, when $\mathcal T_1$ is analyzed to find its best split, the procedure for this task will be {\em FindBestAnchoredSplit}, called directly, instead of passing through {\em FindBestUnanchoredSplit}. So, in our learning model, all splits are binary. Given a node, the `lefthand' outgoing edge is labeled with the chosen $\langle X\rangle p$, or just $p$, when the split is local, whereas the corresponding `righthand' edge is labeled with $[X]\neg p$ (or just $\neg p$); also, the node is labeled with a new reference interval if its corresponding dataset is unanchored. After a split, every $T\in\mathcal T_1$ (the existential dataset, which is now certainly anchored) is associated with a new \emph{witnessing} interval: in fact, those instances satisfy $\langle X\rangle p$ on $[x,y]$, and, for each one of them, there is a possibly distinct witness. Witnesses are assigned by the function {\em Split}; while the witnessing interval of an instance may change during the process, its reference interval is set only once. 

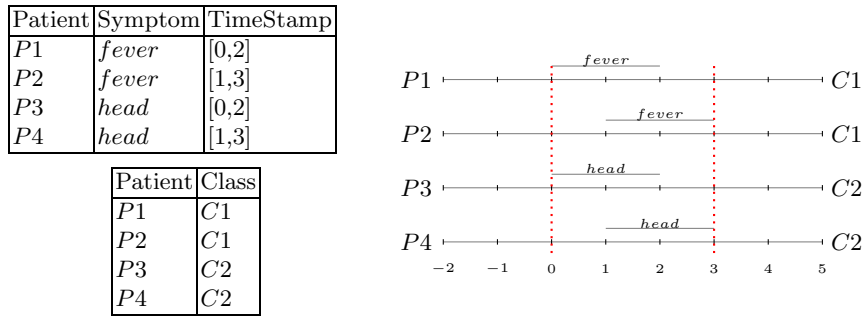
\begin{figure}[t]
\begin{center}
\begin{tikzpicture}[scale=.72]

\node(original1) at (2,-4) {
\begin{tabular}{|l|l|l|}
\hline
Patient & Symptom & TimeStamp \\
\hline
%febbri
$P1$ & $fever$ & [0,2] \\
$P2$ & $fever$ & [1,3] \\
%mal di testa
$P3$ & $head$ & [0,2] \\
$P4$ & $head$ & [1,3] \\
\hline
\end{tabular}
};
\node(original2) at (2.25,-7) {
\begin{tabular}{|l|l|}
\hline
Patient & Class \\
\hline
%febbri
$P1$ & $C1$ \\
$P2$ & $C1$ \\
$P3$ & $C2$ \\
$P4$ & $C2$ \\
\hline
\end{tabular}
};
\foreach \x in {7, 8, 9, 10, 11, 12, 13, 14}
    \draw (\x cm, -39.5mm) -- (\x cm, -40.5mm);
\foreach \x in {7, 8, 9, 10, 11, 12, 13, 14}
    \draw (\x cm, -49.5mm) -- (\x cm, -50.5mm);
\foreach \x in {7, 8, 9, 10, 11, 12, 13, 14}
    \draw (\x cm, -59.5mm) -- (\x cm, -60.5mm);
\foreach \x in {7, 8, 9, 10, 11, 12, 13, 14}
    \draw (\x cm, -69.5mm) -- (\x cm, -70.5mm);
\draw[black, opacity=0.5,thin]  (7,-4) -- (14,-4);
\draw[black, opacity=0.5,thin]  (7,-5) -- (14,-5);
\draw[black, opacity=0.5,thin]  (7,-6) -- (14,-6);
\draw[black, opacity=0.5,thin]  (7,-7) -- (14,-7);
\node(p1t) at (6.5,-4){$P1$};
\node(p2t) at (6.5,-5){$P2$};
\node(p3t) at (6.5,-6){$P3$};
\node(p4t) at (6.5,-7){$P4$};
%eventi
\draw[black, opacity=0.5,thin]  (9,-3.75) -- (11,-3.75);
\node[draw=none] (fever1) at (10.0,-3.65) {\begin{tiny}$fever$\end{tiny}};
%\draw[black, opacity=0.5,thin]  (9,-3.5) -- (11,-3.5);
%\node[draw=none] (head1) at (10,-3.37) {\begin{tiny}$head$\end{tiny}};

\draw[black, opacity=0.5,thin]  (10,-4.75) -- (12,-4.75);
\node[draw=none] (fever2) at (11,-4.65) {\begin{tiny}$fever$\end{tiny}};
%\draw[black, opacity=0.5,thin]  (10,-4.5) -- (12,-4.5);
%\node[draw=none] (head2) at (11,-4.37) {\begin{tiny}$head$\end{tiny}};

\draw[black, opacity=0.5,thin]  (9,-5.75) -- (11,-5.75);
%\node[draw=none] (fever3) at (11,-5.65) {\begin{tiny}$fever$\end{tiny}};
%\draw[black, opacity=0.5,thin]  (9,-5.5) -- (11,-5.5);
\node[draw=none] (fever1) at (10.0,-5.65) {\begin{tiny}$head$\end{tiny}};

\draw[black, opacity=0.5,thin]  (10,-6.75) -- (12,-6.75);
\node[draw=none] (head1) at (11,-6.65) {\begin{tiny}$head$\end{tiny}};
%\node[draw=none] (fever2) at (11,-4.65) {\begin{tiny}$fever$\end{tiny}};

\draw[red,dotted,thick] (9,-3.75) -- (9, -7.25);
\draw[red,dotted,thick] (12,-3.75) -- (12, -7.25);

\node(p1t) at (9, -7.5){\tiny$0$};
\node(p1t) at (10, -7.5){\tiny$1$};
\node(p1t) at (11, -7.5){\tiny$2$};
\node(p1t) at (12, -7.5){\tiny$3$};

\node(p1t) at (7, -7.5){\tiny$-2$};
\node(p1t) at (8, -7.5){\tiny$-1$};
\node(p1t) at (13, -7.5){\tiny$4$};
\node(p1t) at (14, -7.5){\tiny$5$};

\node(p1t) at (14.45,-4){$C1$};
\node(p2t) at (14.45,-5){$C1$};
\node(p3t) at (14.45,-6){$C2$};
\node(p4t) at (14.45,-7){$C2$};

%\draw[->,](middle) -- (sx);
%\draw[->](middle) -- (dx);
\end{tikzpicture}
\end{center} 
\caption{Example of a problematic dataset.}\label{fig:exampleExtended}
\end{figure}

Consider, now, the function {\em AssignReferenceInterval} and the example shown in Figure \ref{fig:exampleExtended}. As can be seen, neglecting the temporal dimension, one may classify the instances with just a single split based on the presence of the symptom fever (or headache). On the contrary, given the temporal dataset with domain $\mathbb{D} = \{0,1,2,3\}$ it is not possible discriminate the classes within a single step. A natural solution consists of augmenting $\mathbb{D}$ in such a way to simulate the behaviour of an infinite domain model. In our example, it suffices to consider $\mathbb{D} = \{-2,-1,0,1,2,3,4,5\}$, so that a single split may be based on the rule: $\langle L \rangle fever \rightarrow C1$, otherwise $C2$ holding on $[-2,-1]$  (or, equivalently, its inverse formulation on $[4,5]$). Thus, the function {\em AssignReferenceIntervals}, while searching all possible reference intervals, takes into consideration two extra points at each side of the domain. Although it is possible to obtain a similar result by adding less than four points (in our example, -2 and -1 suffice), this is no longer true if we include the possibility that Temporal ID3 is called on a {\em subset} of HS modalities, for example, for computational efficiency reasons. Adding four points, on the other hand, guarantees that the most discriminative split can always be found. 

\iffalse
\begin{itemize}
    \item NON E' VERO, ESTENDENDO DI UN SOLO PUNTO A SINISTRA ED UN SOLO PUNTO A DESTRA POSSO FARE LA STESSA COSA CON L'OPERATORE DURING.
    \item E' VERO IL CLAIM CON SOLO QUEI DUE PUNTI? SONO NECESSARI QUATTRO? E SE AVESSIMO PIU' DI UN INTERVALLO PER ISTANZA? 
    \item E METTO SOLO 2 PUNTI A SINISTRA NON POSSO PIU' FARE IL DURING.
    \item IDEA: filosoficamente, dato un dominio finito, diciamo 0 \dots N, ha piu' senso estendere i punti a destra. Infatti, se prendiamo il caso medico, il punto 0 potrebbe corrispondere al giorno in cui ho iniziato il monitoraggio dei pazienti, dunque e' un'origine del tempo, non posso guardare all'indietro. Per contro, anche se il mio dominio arriva ad N, posso estenderlo a N+2 (o, equivalentemente a N+50, N+1000, \dots), in quanto il tempo comunque continua dopo la mia ultima registrazione. Semplicemente non avro' osservazioni su tale parte estesa del dominio, in quanto corrisponde ad un periodo futuro che deve ancora manifestarsi. Si osservi che estendendo con 2 punti a destra posso utilizzare l'operatore inverso a L, e non e' piu' applicabile invece il during.
\end{itemize}
\fi

%For example, $T_1,T_2$ are in $\mathcal T_1$ because %satisfy $\meets p$ on $[x,y]$, it could be the case that  $T_1,[y,z_1]\mmodels p$ and $T_2,[y,z_2]\mmodels p$ with $z_1\neq z_2$. So, $T_1$ and $T_2$ would have two different witnessing intervals, namely $[y,z_1]$ and $[y,z_2]$.

\medskip

\noindent{\bf Analysis.} We now analyze the computational complexity of Temporal ID3. To this end, we first compute the cost of finding the best splitting. Since the cardinality of the domain of each timeline is $N$, there are $O(N^2)$ possible intervals. This means that, fixed a propositional letter and a relation $R_X$, computing $\mathcal T_1$ and $\mathcal T_2$ costs $O(nN^2)$, where $n$ is the number of timelines. Therefore, the cost of {\em FindBestAnchoredSplit} is obtained by multiplying the cost of a single (tentative) splitting by the number of propositional letters and the number of temporal relations (plus one, to take into account the local splitting), which sums up to $O(13 n N^2|\mathcal{AP}|)$. The cost of {\em FindBestUnanchoredSplit} increases by a factor of $N^2$, as the {\bf for} cycle ranges over all possible intervals, and therefore it becomes  $O(13nN^4|\mathcal{AP}|)$. We can increase the efficiency of the implementation by suitably pre-compute the value of $\langle X\rangle p$ for each temporal relation, each propositional letter, and each interval, thus eliminating a factor of $N^2$ from both costs.

%\medskip

If we consider $\mathcal{AP}$ as fixed, and $N$ as a constant, the cost of finding the best splitting becomes $O(n)$, and, under such (reasonable) assumption, we can analyze the complexity of an execution of {\em Learn} in terms of the number $n$ of timelines. Two cases are particularly interesting. In the worst case, every binary split leads to a very unbalanced partition of the dataset, with $|\mathcal T_1|=1$ and $|\mathcal T_2|=n-1$ (or the other way around). The recurrence that describes such a situation is:

    $$t(n)=t(n-1)+O(n),$$

\noindent which can be immediately solved to obtain $t(n)=O(n^2).$
However, computing the worst case has only a theoretical value; we can reasonably expect Temporal ID3 to behave like a randomized divide-and-conquer algorithm, and its computational complexity to tend towards the average case. In the average case, every binary split leads to a non-unbalanced partition, but we cannot foresee the relative cardinality of each side of the partition. Assuming that every partition is equally probable, the recurrence that describes this situation is:

    $$t(n)=\frac{1}{n}\sum\limits_{k=1}^{n}(t(k)+t(n-k))+O(n).$$

    \noindent We want to prove that $t(n)=O(n\log(n))$. To this end, we first prove a useful bound for the expression $\sum\limits_{k=1}^n k\log(k)$, as follows:

    \[\arraycolsep=1.4pt\def\arraystretch{1.5}
  \begin{array}{llll}
    \sum\limits_{k=1}^{n}(k \log(k)) & =  &  \sum\limits_{k=1}^{\lceil \frac{n}{2}\rceil-1}(k \log(k)) + \sum\limits_{k=\lceil \frac{n}{2}\rceil}^{n}(k \log(k))\\
                                      &\le &  \sum\limits_{k=1}^{\lceil \frac{n}{2}\rceil-1}(k \log(\frac{n}{2})) + \sum\limits_{k=\lceil \frac{n}{2}\rceil}^{n}(k \log(n))\\
                                      & =  &  (\log(n)-1)\sum\limits_{k=1}^{\lceil \frac{n}{2}\rceil-1} k + \log(n)\sum\limits_{k=\lceil \frac{n}{2}\rceil}^{n} k\\
                                      & =  &  \log(n)\sum\limits_{k=1}^{n} k - \sum\limits_{k=1}^{\lceil \frac{n}{2}\rceil-1} k\\
                                      &\le &  \frac{1}{2} \log(n) n (n+1)-\frac{1}{2}\frac{n}{2}(\frac{n}{2}+1)\\
                                      & =  &  \frac{1}{2} (n^2 \log(n)+n \log(n))-\frac{1}{8} n^2-\frac{1}{4} n.
    \end{array}
\]

\noindent Now, we prove, by induction, that $t(n) \le a n\log(n) + b$ for some positive constants $a,b$, as follows:
    
    \[\arraycolsep=1.2pt\def\arraystretch{1.5}
    \begin{array}{llll}
    t(n)&=  &\frac{1}{n}\sum\limits_{k=1}^{n}(t(k)+t(n-k))+O(n)&\\
        &=  &\frac{2}{n}\sum\limits_{k=1}^{n}t(k)+O(n) &\\
        &\le&\frac{2}{n}\sum\limits_{k=1}^{n}(a k \log(k)+b)+O(n)& \mbox{inductive hypothesis} \\
    & = &\frac{2}{n}\sum\limits_{k=1}^{n}(a k \log(k))+\frac{2}{n}\sum\limits_{k=1}^{n} b+ O(n)\\
        &=  &\frac{2a}{n}\sum\limits_{k=1}^{n}(k \log(k))+2 b+O(n)\\
        &\le&\frac{2 a}{n}(\frac{1}{2} (n^2 \log(n)+n \log(n))-\frac{1}{8} n^2-\frac{1}{4} n)\\
        & &\,\,+2 b+O(n) & \mbox{proved above} \\
        &=& a n\log(n)+2a \log(n)-\frac{an}{4}-\frac{a}{2}\\
        & &\,\,+2b+O(n)&\\
        &\le& a n \log(n) + b. & \mbox{if }\frac{an}{4}\ge 2a \mbox{log}(n)-\frac{a}{2}+b+O(n).    
    \end{array}
    \]

%\noindent for a large enough value of $a$ such that

%$$\frac{an}{4}\ge 2a \mbox{log}(n)-\frac{a}{2}+b+O(n).$$

%\noindent Finally, we plug in our bound, to obtain:
%\[\arraycolsep=1.4pt\def\arraystretch{1.5}
%    \begin{array}{llll}
%    t(n)&
%    
%    \end{array}
%\]

\begin{figure}[t]
\begin{center}
\begin{tikzpicture}[scale=0.9]
\node(root) at (5,5){$[0,1]$};
\node(left) at (2.5,3){};
\node(right) at (7.5,3){};
\node(leftleft) at (0,1){};
\node(leftright) at (5,1){};
\node(first) at (0,0.5) {$C2$};
\node(second) at (5,0.5) {$C1$};
\node(third) at (7.5,2.5) {$C2$};
\path[-] (root) edge[sloped,above ] node{$\later fever$} (left);
\path[-] (root) edge[sloped, above ] node{$[L]\neg fever$} (right);
\path[-] (left) edge[sloped, above] node{$\Toverlaps head$} (leftleft);
\path[-] (left) edge[sloped, above] node{$[\overline O]\neg head$} (leftright);
\end{tikzpicture}
\end{center}\caption{A decision tree learned by Temporal ID3 on the example in Fig.~\ref{fig:example}.}\label{fig:tree}
\end{figure}
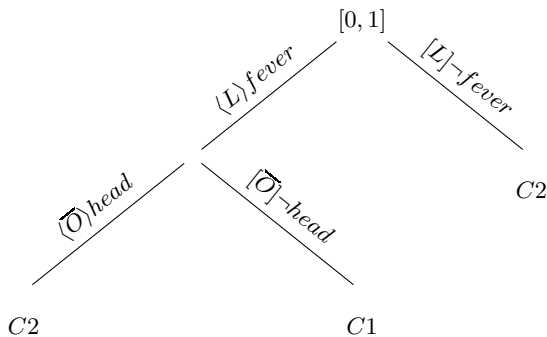

\noindent{\bf Example of execution.} Consider our initial example of Fig.~\ref{fig:example}, with four timelines distributed over two classes. Since this is a toy example, there are many different combination of intervals, relations, and propositional letters that give the same information gain. Fig.~\ref{fig:tree} gives one possible outcome, which seems to indicate that, looking at the entire history, the class $C2$ is characterized by presenting headache and overlapping fever, or no fever at all.

There are several running parameters that can be modulated for an execution of Temporal ID3, and further analysis is required to understand how they influence the final result, and, particularly, the properties of the resulting classifier. The most important ones are: \begin{inparaenum}[\it (i)] \item how to behave in case of two splits with the same information gain; \item how to behave in case of more than one possible witness interval for a given timeline; \item how to behave in case of more than one optimal reference interval for a given unanchored temporal dataset. 
\end{inparaenum} If we allow, in all such cases, a random choice, the resulting learning algorithm is not deterministic anymore, and different executions may result in different decision trees. This is a typical situation in machine learning (e.g., in algorithms such as {\em k-means clustering}, or {\em random forest}), that involves some experience in order to meaningfully assess the results.

\section{Conclusions}

Classical decision trees, which are a popular class of learning algorithms, are designed to interpret categorical and numerical attributes. In decision trees, every node can be seen as a propositional letter; therefore, a decision tree can be seen as a structured set of propositional logic rules, the right-hand part of which is a class. Since classifying based on the static aspects of data is not always adequate, and since decision tree learning cannot deal with temporal knowledge in an explicit manner, we considered the problem of learning a classification model capable to combine propositional knowledge with qualitative temporal information. Towards its solution, we showed how temporal data can be prepared in a optimal way for a temporal decision tree to be learned and presented a generalization of the classical decision tree learning algorithm ID3 that is able to split the dataset based on temporal, instead of static, information, using the well-known temporal logic HS. Future work include testing our method on real data, improving the capabilities of Temporal ID3 by enriching the underlying language, and studying the effect of different pruning and stopping conditions. Moreover, it would be interesting to study adapting ID3 to other logical languages, although this may require re-designing some key elements, such as the representation of temporal datasets, or the process that underlies the splitting algorithm. 

%\medskip

Machine learning is generically focused on a non-logical approach to knowledge representation. However, when learning should take into account temporal aspects of data, a logical approach can be associated to classical methods, and besides decision tree learning, interval temporal logics has been already proposed as a possible tool, for example, for temporal rules extraction~\cite{DBLP:conf/time/MonicaFMMS17}. Focusing these approaches on fragments of interval temporal logics whose satisfiability problem is decidable (and tractable) may result into an integrated systems that pairs induction and deduction of formulas, intelligent elimination of redundant rules, and automatic verification of inducted knowledge against formal requirement. Also, using a logical approach in learning may require non-standard semantics for logical formulas (e.g., fuzzy semantics, or multi-valued propositional semantics); these, in turn, pose original and interesting questions on the theoretical side concerning the computational properties of the problems associated with these logics (i.e., satisfiability), generating, {\em de facto}, a cross-feeding effect on the two fields. 

%Among all classification models, decision trees are still very popular. Their success is due to the computational efficiency of the learning process, the interpretability of tree models, and their wide versatility. 

%Real data testing is not an easy task: data that contain temporal information, such as medical data (as per our example) are usually not prepared for timeline analysis. Therefore, a non-trivial algorithmic effort must be devoted to data preparation, with particular attention to correct and meaningful labeling of intervals.

\bibliographystyle{splncs04.bst}
\bibliography{biblio}

\begin{thebibliography}{10}
\providecommand{\url}[1]{\texttt{#1}}
\providecommand{\urlprefix}{URL }
\providecommand{\doi}[1]{https://doi.org/#1}

\bibitem{allen83}
Allen, J.F.: Maintaining knowledge about temporal intervals. {Communications}
  of the ACM  \textbf{26}(11),  832--843 (1983)

\bibitem{DBLP:conf/iwcs/Alluhaibi15}
Alluhaibi, R.: Simple interval temporal logic for natural language assertion
  descriptions. In: Proc. of the 11th International Conference on Computational
  Semantics ({IWCS}). pp. 283--293 (2015)

\bibitem{DBLP:journals/iandc/Angluin87}
Angluin, D.: Learning regular sets from queries and counterexamples.
  Information and Computation  \textbf{75}(2),  87--106 (1987)

\bibitem{Antipov2011}
Antipov, S., Fomina, M.: A method for compiling general concepts with the use
  of temporal decision trees. Scientific and Technical Information Processing
  \textbf{38}(6),  409--419 (2011)

\bibitem{DBLP:conf/cicling/Baeza-Yates04}
Baeza{-}Yates, R.: Challenges in the interaction of information retrieval and
  natural language processing. In: Proc. of the 5th International on
  Computational Linguistics and Intelligent Text Processing ({CICLing}). pp.
  445--456 (2004)

\bibitem{DBLP:conf/formats/BartocciBS14}
Bartocci, E., Bortolussi, L., Sanguinetti, G.: Data-driven statistical learning
  of temporal logic properties. In: Proc. of the 12th International Conference
  on Formal Modeling and Analysis of Timed Systems. pp. 23--37 (2014)

\bibitem{Blockeel:1998:TIF:1643275.1643308}
Blockeel, H., De~Raedt, L.: Top-down induction of first-order logical decision
  trees. Artificial Intelligence  \textbf{101}(1-2),  285--297 (1998)

\bibitem{DBLP:conf/hybrid/BombaraVPYB16}
Bombara, G., Vasile, C., Penedo, F., Yasuoka, H., Belta, C.: A decision tree
  approach to data classification using signal temporal logic. In: Proc. of the
  19th International Conference on Hybrid Systems: Computation and Control. pp.
  1--10 (2016)

\bibitem{cart84}
Breiman, L., Friedman, J., Olshen, R., Stone, C.: Classification and regression
  trees. Wadsworth and Brooks, Monterey, CA (1984)

\bibitem{sosym}
Bresolin, D., {Della Monica}, D., Goranko, V., Montanari, A., Sciavicco, G.:
  Metric propositional neighborhood logics on natural numbers. Software and
  System Modeling  \textbf{12}(2),  245--264 (2013)

\bibitem{DBLP:journals/tcs/BresolinMMSS14}
Bresolin, D., Monica, D.D., Montanari, A., Sala, P., Sciavicco, G.: Interval
  temporal logics over strongly discrete linear orders: Expressiveness and
  complexity. Theoretical Computer Science  \textbf{560},  269--291 (2014)

\bibitem{ijfcs2012}
Bresolin, D., Sala, P., Sciavicco, G.: On begins, meets, and before.
  International Journal of Foundations of Computer Science  \textbf{23}(3),
  559--583 (2012)

\bibitem{DBLP:conf/icist/BrunelloMMS18}
Brunello, A., Marzano, E., Montanari, A., Sciavicco, G.: {J48S:} {A} sequence
  classification approach to text analysis based on decision trees. In: Proc.
  of the 24th International Conference on Information and Software
  Technologies, {(ICIST)}. pp. 240--256 (2018)

\bibitem{DBLP:conf/isola/BufoBSBLB14}
Bufo, S., Bartocci, E., Sanguinetti, G., Borelli, M., Lucangelo, U.,
  Bortolussi, L.: Temporal logic based monitoring of assisted ventilation in
  intensive care patients. In: Proc. of the 6th International Symposium on
  Leveraging Applications of Formal Methods, Verification and Validation. pp.
  391--403 (2014)

\bibitem{cheng2007discriminative}
Cheng, H., Yan, X., Han, J., Hsu, C.: Discriminative frequent pattern analysis
  for effective classification. In: Proc. of the 23rd International Conference
  on Data Engineering {(ICDE)}. pp. 716--725 (2007)

\bibitem{DBLP:journals/jair/ConsolePD03}
Console, L., Picardi, C., Dupr{\'{e}}, D.: Temporal decision trees: Model-based
  diagnosis of dynamic systems on-board. J. Artif. Intell. Res.  \textbf{19},
  469--512 (2003)

\bibitem{DBLP:conf/time/MonicaFMMS17}
{Della Monica}, D., de~Frutos{-}Escrig, D., Montanari, A., Murano, A.,
  Sciavicco, G.: Evaluation of temporal datasets via interval temporal logic
  model checking. In: Proc. of the 24th International Symposium on Temporal
  Representation and Reasoning ({TIME}). pp. 11:1--11:18 (2017)

\bibitem{fan2008direct}
Fan, W., Zhang, K., Cheng, H., Gao, J., Yan, X., Han, J., Yu, P., Verscheure,
  O.: Direct mining of discriminative and essential frequent patterns via
  model-based search tree. In: Proc. of the 14th ACM SIGKDD {I}nternational
  {C}onference on {K}nowledge {D}iscovery and {D}ata {M}ining. pp. 230--238
  (2008)

\bibitem{fournier2014vgen}
Fournier-Viger, P., Gomariz, A., {\v{S}}ebek, M., Hlosta, M.: {VGEN}: fast
  vertical mining of sequential generator patterns. In: Proc. of the 16th
  International Conference on Data Warehousing and Knowledge Discovery
  {(DaWaK)}. pp. 476--488 (2014)

\bibitem{HalpernS91}
Halpern, J., Shoham, Y.: A propositional modal logic of time intervals.
  {Journal} of the ACM  \textbf{38}(4),  935--962 (1991)

\bibitem{KarimiHamilton}
Karimi, K., Hamilton, H.J.: Temporal rules and temporal decision trees: {A}
  {C4.5} approach. Tech. Rep. CS-2001-02, Department of Computer Science,
  University of Regina (2001)

\bibitem{DBLP:conf/ismis/LinO00}
Lin, W., Orgun, M.: Temporal data mining using hidden periodicity analysis. In:
  Proc. of the 12th International Symposium on Foundations of Intelligent
  Systems ({ISMIS}). pp. 49--58 (2000)

\bibitem{Mballo:2005}
Mballo, C., Diday, E.: Decision trees on interval valued variables. Symbolic
  Data Analysis  \textbf{3}(1),  8--18 (2005)

\bibitem{DBLP:conf/hybrid/NguyenKJDBJ17}
Nguyen, L., Kapinski, J., Jin, X., Deshmukh, J., Butts, K., Johnson, T.:
  Abnormal data classification using time-frequency temporal logic. In: Proc.
  of the 20th International Conference on Hybrid Systems: Computation and
  Control. pp. 237--242 (2017)

\bibitem{pratt}
Pratt-Hartmann, I.: Temporal prepositions and their logic. Artificial
  Intelligence  \textbf{166}(1--2),  1--36 (2005)

\bibitem{DBLP:journals/ml/Quinlan86}
Quinlan, J.: Induction of decision trees. Machine Learning  \textbf{1},
  81--106 (1986)

\bibitem{Quinlan:1987:SDT:50007.50008}
Quinlan, J.: Simplifying decision trees. International Journal of
  Human-Computer Studies  \textbf{51}(2),  497--510 (1999)

\bibitem{Rajan:2006:ART:1218776.1218799}
Rajan, A.: Automated requirements-based test case generation. SIGSOFT Software
  Engeneering Notes  \textbf{31}(6), ~1--2 (2006)

\bibitem{8465421}
Vagin, V., Morosin, O., Fomina, M., Antipov, S.: Temporal decision trees in
  diagnostics systems. In: 2018 International Conference on Advances in Big
  Data, Computing and Data Communication Systems (icABCD). pp. 1--10 (2018)

\bibitem{witten2016data}
Witten, I., Frank, E., Hall, M., Pal, C.: Data Mining: Practical Machine
  Learning Tools and Techniques. Morgan Kaufmann (2016)

\end{thebibliography}

\end{document}